\documentclass[aps,prl,showpacs,twocolumn]{revtex4}

\usepackage{amsmath,amssymb,graphicx}

\begin{document}

%%%%%%%%%%%%%%%%%%%%%%%%%%%%%%%%%%%%%%%%%%%%%%%%%%%%%%%%%%%%%%%%%%%%%%%%%%%%%%%%%%%%%%%%%%%%%%%%%%%%%%%%%%%%%%%%%%%%%%%%%%%%
\noindent{\bf Comment on \textquotedblleft Semiconducting Layered Blue Phosphorus: A Computational Study\textquotedblright}\\

%%%%%%%%%%%%%%%%%%%%%%%%%%%%%%%%%%%%%%%%%%%%%%%%%%%%%%%%%%%%%%%%%%%%%%%%%%%%%%%%%%%%%%%%%%%%%%%%%%%%%%%%%%%%%%%%%%%%%%%%%%%%%%
\noindent Using results drawn from the pages of PRL\cite{Zhu2014},
the authors comment on \textquotedblleft Semiconducting Layered 
Blue Phosphorus: A Computational Study\textquotedblright.
In recent letter\cite{Zhu2014} unknown phase of phosphorus
with high stability and a wide fundamental gap was proposed 
by Zhu, and Tom\'anek. The first half of the comment questioned
elements of phosphorus blue structure, while the second half devoted
to van der Waals forces between AB stacked blue phosphorus
layers which missed by them.

Converting a mono-layer of black to blue phosphorus, they used
dislocation with constantlocal band angles and changed z lattice
vector direction. This dislocation is not complete method for 
predicting new structure, While predicting a new system is 
available via {\it ab initio} evolutionary algorithm implemented for 
instance in USPEX code
\cite{Uspex1,Uspex2,Uspex3}.  

Using USPEX code accompanied with VASP~\cite{Kresse2007}, We found  black phosphorous as  the most stable structure  
 and after that  the blue phosphorous is stable 
 with the difference energy of  2.62 eV. 
 The interlayer distance, d$_{int}$, in above structure from USPEX prediction for Black and Blue ph. is 5.78 and 3.56 \AA{}
 that is near to our results from ven der Waals calculation in table~\ref{Tab1}.
 Then  in monolayer structure, we found 
 black monolayar is 0.034 eV stable than blue phosphorous monolayer.
 
 \begin{figure}[htbp]
\includegraphics*[width=0.35\textwidth]{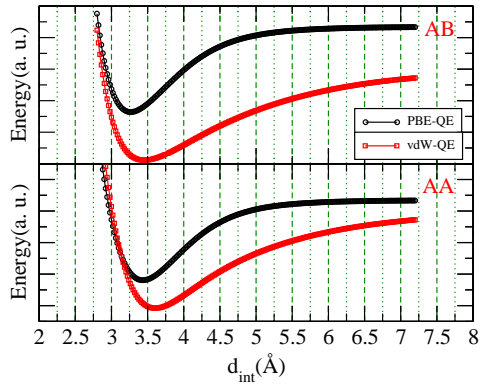}  
\caption{\label{Scheme} Total interaction energy between two layer in AB and AA stacked blue phosphorous  
separated by the zero \AA{} thick vacuum.} 
\end{figure}
Presented here, reveals,
the interlayer distance adopted in the letter are not
in fact the most stable distance. The phosphorus layers, especially
in blue configuration, stacked together by van der 
Waals interactions, like graphite\cite{Likai2014}.

The Brillouin zone of primitive cell for all calculation was 
sampled using 10$\times$10$\times$1~k points in
standard DFT calculation pseudo-potential
methods by  Quantum Espresso (QE)
(ecutwfc=40.0 and ecutrho=500) \cite{QE-2009}.
%We used the pseudopotentials 
%p-pbe-v1.uspp.F.UPF from ref.~\cite{http}.
\begin{table}[htb]
\caption{\label{Tab1}The nearest distance between the adjacent layer (d$_{int}$) interlayer distance for a, zero vacuum, and b, 15 \AA{} vacuum,.
$\Delta$E$^{'}$(Ry*10$^{-2}$) is, $\Delta$E, energy difference between minimum energy level and straight energy level per atom.
Values in parenthesis is related to black phosphorus.}
\begin{ruledtabular}
%\tiny
\begin{tabular}{c| c| c| c| c }
 name                    &  a                 &a                &b                                 &b                           \\
 \hline
                         & d$_{int}$(\AA{})   &$\Delta$E$^{'}$    & d$_{int}$(\AA{})                 & $\Delta$E$^{'}$             \\ 
\hline 
AB-PBE               &  3.26(5.40)        & 1.35(0.20)  &  3.06(5.82)                      & 0.99(0.02)             \\
\hline
AB-vdW-DF                &  3.46(5.62)        & 1.30(0.46)  &  3.12(5.86)                      & 0.69(0.17)             \\
\hline
AA-PBE                &  3.44(5.24)        & 1.26(0.39)  &  4.12(5.86)                      & 0.10(0.03)              \\
\hline
AA-vdW-DF                &  3.60(5.46)        & 1.40(0.59)  &  4.24(5.88)                      & 0.32(0.18)              \\
\end{tabular}
\end{ruledtabular}
\end{table}
\\
\\
Concerning van der Waals (vdW)
interaction via self consistent vdW-DF implemented in
QE code, We found 3.46 and 3.12 \AA{} as 
interlayer distance for two periodic array of AB stacked slab 
separated by a viz. a) zero and b)15 \AA{} thick vacuum
region, respectively.  This distance is completely different
of 5.63 angstrom mentioned in FIG. 1d of Letter \cite{Zhu2014}.
In same calculation for AA stacked, d$_{int}$ is 3.60 and 4.24 \AA{}
(see table \ref{Tab1} and figure \ref{Scheme}).
we repeated this calculation for black phosphorus, so interlayer distance 
reached its minimum in $\simeq$ 5.6 \AA{}, in accordance
with experimental results \cite{Likai2014}.
The difference between minimum energy level and straight energy level 
in black one is around 0.5*10$^{-2}$ Ry, while in blue one is 1.30*10$^{-2}$ Ry.

%%%%%%%%%%%%%%%%%%%%%%%%%%%%%%%%%%%%%%%%%%%%%%%%%%%%%%%%%%%%%%%%%%%%%%%%%%%%%%%%%%%%%%%%%%%%%%%%%%%%%%%%%%%%%%%%%%%%%%%%%%%%%%%
%\noindent {\it Acknowledgement.} 
%This work is funded by the Science Foundation of Ireland (??????),
%by the EU-FP7 (iFOX project????) and by Isfahan University of
%technology. Computational resources have been provided by
%the Trinity Center for High Performance Computing and CMS
%group computing resources.\\

%%%%%%%%%%%%%%%%%%%%%%%%%%%%%%%%%%%%%%%%%%%%%%%%%%%%%%%%%%%%%%%%%%%%%%%%%%%%%%%%%%%%%%%%%%%%%%%%%%%%%%%%%%%%%%%%%%%%%%%%%%%%%%%
\vspace{1mm}

\noindent I.~Abdolhosseini Sarsari$^1$, Z.~Allahyari$^{1,2,3}$, M.~Alaei$^1$ and S.~Sanvito$^4$ \\
$^1$Department of Physics, Isfahan University of Technology, Isfahan, 84156-83111, Iran\\
$^2$Moscow Institute of Physics and Technology, Dolgoprudny, Moscow Region, 141700, Russia\\
$^3$Skolkovo Institute of Science and Technology, Skolkovo Innovation Center, Bldg. 3, Moscow 143026, Russia\\
$^4$School of Physics, AMBER and CRANN, Trinity College, Dublin 2, Ireland\\

%%%%%%%%%%%%%%%%%%%%%%%%%%%%%%%%%%%%%%%%%%%%%%%%%%%%%%%%%%%%%%%%%%%%%%%%%%%%%%%%%%%%%%%%%%%%%%%%%%%%%%%%%%%%%%%%%%%%%%%%%%%%%%%%

\end{document}